\documentclass[aps,pra,reprint,nofootinbib,showpacs,superscriptaddress]{revtex4-1}
\usepackage{amssymb,amsmath,latexsym}
\usepackage{setspace}
\usepackage{braket}
\usepackage{graphicx}
\usepackage{dcolumn}
\usepackage{bm}
\usepackage{hyperref}
\usepackage{amsfonts}
\usepackage{amsmath,amssymb}
\usepackage{feynmp}
\usepackage{amsmath,amsthm}
\usepackage{amssymb}
\usepackage{graphicx,color}
\usepackage{subfigure}
\usepackage{epstopdf}
\usepackage{color}
\usepackage{natbib}
\usepackage[strict]{changepage}
\usepackage{epsfig}
\usepackage{fancybox}
\usepackage{listings}
\usepackage{gensymb}
\usepackage{url}
\usepackage{verbatim}
\newcommand{\eval}[2][\right]{\relax \ifx#1\right\relax \left.\fi#2#1\rvert}

 \def\be{\begin{equation}}
 \def\ee{\end{equation}}
 \def\bes{\begin{eqnarray}}
 \def\ees{\end{eqnarray}}

 \def\2{\frac{1}{2}}
 \def\4{\frac{1}{4}}

\begin{document}

\noindent\fbox{%
    \parbox{\textwidth}{%
        This manuscript has been authored by UT-Battelle, LLC under Contract No. DE-AC05-00OR22725 
with the U.S. Department of Energy. The United States Government retains and the publisher, 
by accepting the article for publication, acknowledges that the United States Government retains 
a non-exclusive, paid-up, irrevocable, world-wide license to publish or reproduce the published 
form of this manuscript, or allow others to do so, for United States Government purposes. The 
Department of Energy will provide public access to these results of federally sponsored research
in accordance with the DOE Public Access Plan (http://energy.gov/downloads/doe-public-access-plan).
    }%
}
\clearpage

\title{Toroidal nano-traps for cold polar molecules}
\author{Marouane Salhi} \email{msalhi@vols.utk.edu}
\affiliation{Department of Physics and Astronomy, The University of Tennessee, Knoxville, TN 37996-1200, USA}
\author{Ali Passian} \email{passianan@ornl.gov}
\affiliation{Computational Sciences and Engineering Division, Oak Ridge National Laboratory, Oak Ridge, TN
37831-6418, USA} 
\author{George Siopsis} \email{siopsis@tennessee.edu}
\affiliation{Department of Physics and Astronomy, The University of Tennessee, Knoxville, TN 37996-1200, USA}
\date{\today}
\begin{abstract}  
Electronic excitations in metallic nanoparticles in the optical regime that have been of great importance in surface enhanced spectroscopy and emerging applications of molecular plasmonics, due to control and confinement of electromagnetic energy, may also be of potential to control the motion of nanoparticles and molecules. Here, we propose a concept for trapping polarizable particles and molecules using toroidal metallic nanoparticles. Specifically, gold nanorings are investigated for their scattering properties and field distribution to computationally show that the response of these optically resonant particles to incident photons permit the formation of a nanoscale trap when proper aspect ratio, photon wavelength and polarization are considered.  However, interestingly the resonant plasmonic response of the nanoring is shown to be detrimental to the trap formation. 
The results are in good agreement with analytic calculations in the quasi-static limit within the first-order perturbation of the scalar electric potential.  
The possibility of extending the single nanoring trapping properties to two-dimensional arrays of nanorings is suggested by obtaining the field distribution of nanoring dimers and trimers.
\end{abstract}
\maketitle
\section{Introduction}
The development of Penning~\cite{Dehmelt1967} and Paul traps~\cite{Paul1953} led to unprecedented measurements of fundamental properties of charged particles by isolating individual electrons and ions~\cite{Jr1987}. Singling out and having the ability to probe individual ions allows laser manipulation of their quantum states to form qubits~\cite{Harty2014,Kim2014}, which is capitalized upon in the so called trapped ion quantum computer~\cite{Kielpinski2002}. Precision measurements benefit from trapping or even slowing down the quantum systems, as reported by Veldhoven, \emph{et al.}, who carried out high resolution spectroscopy by slowing down a pulsed $^{15}$ND$_3$ molecular beam in a Stark-decelerator~\cite{Veldhoven2004}. Deceleration can further help introducing  cold polar molecules into traps or storage rings~\cite{VandeMeerakker2008}. 
For a comprehensive account on the importance of polar molecules and the selection of those suited for deceleration and trapping experiments, see the recent review by van de Meerakker, \emph{et al.}~\cite{VandeMeerakker2012}. 
Similarly, the development of optical trapping~\cite{Marago2013} of small particles, capitalizing on force gradients associated with radiation pressure~\cite{Ashkin1970} exerted by tightly focused laser beams, led to the development of optical tweezers~\cite{Ashkin1986}. Optical trapping in the tweezer configuration has produced significant volumes of data including those related to measurements of micro-rheological properties of aerosol particles~\cite{Power2014}, optical binding of trapped particles~\cite{Bowman2013}, and DNA and cell studies~\cite{Bustamante2008}.    
\begin{figure}
\centering
\includegraphics[width=9.5cm]{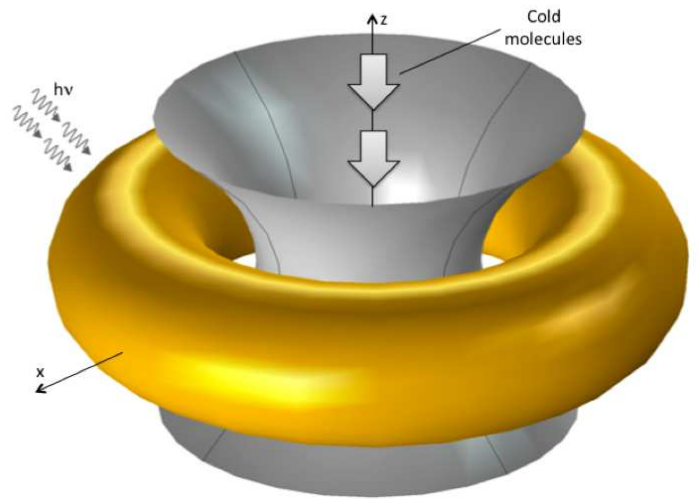}
\caption{\label{fig:1}The concept of a nanoring based toroidal trap. Cold polar molecules near the gray shaded surface approaching the central region may be trapped within a nanometer scale volume due to the potential barrier established by the optical response of the torus to an incoming beam of photons with proper polarization, wavelength, and amplitude modulation.}
\end{figure} 

The ability to manipulate the spatial distribution and time dependence of the electromagnetic fields interacting with charged particles or neutral molecules of finite dipole moments is of 
prime importance in the further development of methods to contain and control single or ensembles of molecules~\cite{Lemeshko2013}. 
Local confinement and enhancement of electromagnetic fields, per recent advances in nano-optics and plasmonics~\cite{Schuller2010},  will undoubtedly prove important in the further development of systems that can exert electromagnetically induced spatial constraints on molecular and particulate populations.  
For applications in quantum information processing, it is important to have the ability to trap arrays of particles~\cite{DeMille2002}. In particular, the development of arrays of electrodynamic nano-traps, as proposed here, may offer an interesting perspective in storing and manipulating well separated but compact set of molecules.
Advanced nano-fabrication~\cite{Gates2005} may be employed to  generate a plethora of nanostructures~\cite{Xia2003} that can alter the electromagnetic fields to create specific force fields within which molecules may reside in slow motion. 
A variety of trap configurations and techniques have been proposed and designed, in particular for cold polar molecules~\cite{Bethlem2002a,DeMille2004a,Junglen2004,DeMille2004,Meerakker2005,Rieger2005,VanVeldhoven2005,Xia2005,Bethlem2006,VanVeldhoven2006,Kleinert2007a,Sawyer2007,Schnell2007,Micheli2007}. 
For on chip applications, electrostatic surface trapping of cold polar molecules have been proposed using various electrode configurations including a single charged circular wire~\cite{Hui2007}, parallel charged wires~\cite{Xia2005a}, and  two charged rings above a grounded conductor plate embedded in an insulator~\cite{Li2013}. 

In this article, we propose and investigate the concept of sub-wavelength electromagnetic traps in the form of metallic toroidal structures (micro- and nano-rings), shown in Fig.~\ref{fig:1}. Throughout this work, the term trap is defined to describe the electromagnetic environment in which a particle assumes a dynamic state with an eigenenergy  spectrum insufficient to overcome a potential barrier. Similarly, nano-trap refers to a physical arrangement in which appropriate particles experience a potential barrier with a spatial extension in the order of nanometers. We propose to achieve a nano-trap with toroidal geometry using noble metal nano-rings, which upon interaction with long wavelength linearly polarized photons produce a three-dimensional trapping potential barrier.  The barrier shape is shown to be further controlled by use of circularly polarized photons.    
We envision the toroidal nano-trap can be used to capture polarizable dielectric nanoparticles, similar to optical tweezers~\cite{Juan2011},   as well as cold polar molecules, assuming that the particles and molecules have been through a cooling process to reach a temperature of a few mK using an appropriate deceleration mechanism, for example by being transmitted through a Stark decelerator in the case of molecular beams as  reported by Bethlem, \emph{et al.}~\cite{Bethlem1999}. 
The studied structures are topologically distinct from the more commonly used geometries of a sphere, cylinder, or cube~\cite{Sosa2003}: tori possess  a distinct  electromagnetic field distribution that is here shown to allow for the formation of a trap for particles and polar molecules.  
As a comparison,  a quadrupole trap typically consists of a ring electrode with
two end-caps. Here, in addition to the length scale difference, the proposed concept achieves an efficient trap in a configuration without end-caps. 
 The designation \emph{toroidal nanotrap} thus emphasizes the geometry of the structure employed to create a trapping region rather than the form of the trapping field. The actual trapping volume will be shown to take a rather spheroidal shape.  Furthermore, comparing the proposed concept to quadrupole or hexapole electrode based traps~\cite{Kleinert2007a, Bethlem2006}, it will be shown that, while the field lines in both cases may bear certain resemblance in  the central region of the trap, the toroidal optically induced trap is solely due to the dielectric response of the metal nanoring.
However, the advantages of optical excitation  as opposed to applying a voltage are manifold: a) optical trapping can be achieved off the chip or remotely without making physical connections, b) optical fields can be provided with high modulation rates, polarization control, wavelength tuning, controllable power and intensity profiles, all of which can help achieving the desired field properties, c) optical fields can be delivered dynamically (as a rough comparison we may consider the method of laser scanning in digital optical disc data storage devices), d) multiple lasers may be employed simultaneously, each with a desired polarization state, but with well defined phase relationship, and e) optical excitation is compatible with direct voltage application and thus in a proper configuration can be used in conjunction with the latter for other potential applications.

The presentation is organized as follows.
We begin by characterizing the optical response of a single torus in Section \ref{sec:2}. In doing so, we will first treat the problem
analytically  in the quasi-static limit (where the field is generated by the scalar potential only) since the trapping properties are investigated in a spectral range where the wavelength of the force inducing photons will be larger than the dimensions of the nano-rings. Computational techniques are then invoked to obtain the time dependent scattering properties.
By defining a trapping volume in Section \ref{sec:3}, we discuss the toroidal trapping energy barrier for $^{15}$ND$_3$, a candidate molecular system for deceleration studies~\cite{Bethlem2002b}.
Discussing the results in Section \ref{sec:4}, we also introduce multi-ring systems or arrays with the potential to control a lattice of polar molecules or particulates -- an arrangement which may be useful in quantum computing and quantum information processing ~\cite{DeMille2002} and in sensing and imaging~\cite{Shipway2000}. Concluding remarks are provided in Section \ref{sec:5}.

\section{Field distribution and scattering properties of toroidal particles}
\label{sec:2}

The potential barrier experienced by a trapped particle is assumed here to be the result of the interaction of the particle with an applied electric field modulated locally by the torus. Since our objective is to describe the concept and potential use of the metal torus as a particle trap, we assume that electromagnetic properties of the trapped particle will not significantly affect the response of the trapping particle. This  yields a decoupling of the problem such that the field of the trapping particle may be obtained while neglecting any modification induced by the radiative or polarizing effects of the trapped particle. 
The solid material domain making up the metallic torus is defined by a local dielectric function $\epsilon$ (and conductivity $\sigma$).
Within the spectral and sub-Kelvin temperature ranges considered, we will safely neglect the temperature dependence of $\epsilon$, which otherwise has been shown to create thermoplasmonic effects in thin films~\cite{Lereu2008a}  and nano-particles~\cite{Lereu2013}.
To obtain the field modification engendered by the presence of the torus, we aim to characterize the scattering properties of gold nano-rings, beginning with a single isolated nano-ring in vacuum. We therefore consider a torus, defined by fixing one of the coordinates $u = u_0$ in the toroidal system $(u,\eta,\phi)$, which is related to Cartesian coordinates  via:
  \bes x = \frac{s\sqrt{u^2 -1}}{u - \cos \eta} \cos\phi, \nonumber
y =  \frac{s\sqrt{u^2 -1}}{u - \cos \eta} \sin\phi,
  z =  \frac{s\sin\eta}{u - \cos\eta}, \ees
where $u\in [1,\infty [$, $\eta, \phi \in [0,2\pi]$, and $s$ is a scale factor. 
 We first consider the response of the nano-particle in the quasi-static regime to an arbitrarily polarized uniform applied field $\bar{E_a} = E_a(\sin\theta, 0, \cos\theta)$, where $\theta$ is measured from the $z$~axis. 
 Introducing 
 a perturbation parameter $\xi$, we 
 expand the scalar potential $\Psi_a$ in the toroidal system as:
\begin{align}
 \Psi_{\text{a}} &= E_a(x\sin \theta + z \cos \theta)  \\
                       &=  -2\sqrt{2}\pi^{-1}sE_a\sqrt{u - \xi\cos\eta} \sum_ {n=0}^\infty {\cal{G}}_n(\bar{r},\theta;\xi),
\end{align}
where 
\begin{multline}
   {\cal{G}}_n(\bar{r},\theta;0) =    \sin\theta \cos\phi \cos n\eta  \left( 1 - \frac{\delta_{n0}}{2} \right)  Q_{n- \frac{1}{2}}^1 ( u)  \\ +  \cos\theta \sin n\eta (1-\delta_{n0}) n Q^0_{n - \frac{1}{2}} (u), 
\end{multline}
with $P_n^m(u)$ and $Q_n^m(u)$ denoting the Legendre functions.
Here, $\xi$ is introduced as a bookkeeping parameter for the perturbative calculations; at the end, we set $\xi =1$. The polarization $\theta$ can be set by a polarization rotator and may be treated as a parameter so that $\theta =0$ generates $\bar{E_a} \parallel  \hat{z}$, while  $\theta =\pi/2$ generates $\bar{E_a} \perp  \hat{z}$. 
The field $\bar{E_a}$ is modified by the response of the torus to produce a total field, which we will study 
in two cases of  polarization consistent with the toroidal symmetry.

We now proceed to obtain the potential distribution in the case of $E_{a\parallel}$ by superimposing $\Psi_a$ and the solution of the Laplace equation. The  potential of the trapping field  $\bar{E}$ then takes the form:
\be \Psi = \mathcal{N} \sum_{n=1}^\infty n \sin n\eta \left[  a_n \frac{Q^0_{n- \frac{1}{2}} (u_0)}{P^0_{n-\frac{1}{2}} (u_0)} P^0_{n-\frac{1}{2}} (u)+ Q^0_{n- \frac{1}{2}} (u) \right], \ee
where $\mathcal{N}(u) = -2\sqrt{2}\pi^{-1}sE_a\sqrt{u},$ and  the coefficient  $a_n$ can be obtained from the continuity of the potential and the $u$-component of $\bar D = \epsilon \bar E$
across the toroidal boundary $u = u_0$. At zeroth perturbative order ($\xi =0$), we obtain $ a_n = (1-\epsilon)/( \epsilon - c_n )$, where 
\be
c_n =   \left. \frac{\frac{d}{du}   \log\left[ \sqrt{u} P^0_{n- \frac{1}{2}} (u) \right]}{ \frac{d}{du}   \log\left[ \sqrt{u} Q^0_{n- \frac{1}{2}} (u) \right]} \right|_{u=u_0}.
\ee
Similarly, in the case of perpendicular $E_\perp$ polarization, specifically $\bar{E}\parallel \hat{x}$, 
the potential distribution of the trapping field is expanded as:
\bes \label{ex}
\Psi &=& \mathcal{N} \cos\phi \sum_{n=0}^\infty \left( 1 - \frac{\delta_{n0}}{2} \right) \cos n\eta \nonumber\\
&&\ \ \ \ \ \ \ \ \times \left[  a_n^1 \frac{Q_{n- \frac{1}{2}}^1 (u_0)}{P_{n-\frac{1}{2}}^1 (u_0)} P_{n-\frac{1}{2}}^1 (u)+ Q_{n- \frac{1}{2}}^1 (u) \right], \ \ \ \  \ees
which at the zeroth perturbative order, produces $a_n^1 = (1-\epsilon)/( \epsilon       - c_n^1 ) $ with
\be 
c_n^1 =   \left. \frac{\frac{d}{du}   \log\left[ \sqrt{u} P_{n- \frac{1}{2}}^1 (u) \right]}{ \frac{d}{du}   \log\left[ \sqrt{u} Q_{n- \frac{1}{2}}^1 (u) \right]} \right|_{u=u_0}.
  \ee
In the quasi-static dipole approximation, the scattering cross section of the ring can be obtained by considering only the dipolar contribution of the induced surface charge on the ring to the scattered field. Using the exact interior field and the exact exterior total field (as opposed to the first order approximate field in Eq.~\eqref{ex}), the dipolar charge density $\sigma_c$ is found to be:
\begin{equation}\label{surfacecharge}\sigma_{c}(\eta,\phi) = (u_0-\sin\eta)^{3/2} \sum_{n=0}^\infty M_n \cos n\eta \cos\phi, \end{equation}
where 
\begin{equation*}
\begin{split}
M_n &= \frac{2\epsilon_0\sqrt2}{\pi}\sqrt{u_0^2-1}  \left[A_n-E_a(1+\delta_{n0})\right]  \\
& \times  \left[ Q^{{1}^{\prime}}_{n-\frac{1}{2}}(u_0)-P^{{1}^{\prime}}_{n-\frac{1}{2}}(u_0)\frac{Q^{1}_{n-\frac{1}{2}}(u_0)}{P^{1}_{n-\frac{1}{2}}(u_0)} \right],
\end{split}
\end{equation*}
and the coefficients $A_n$ will be obtained from the continuity of the displacement field $\bar{D}$, yielding the following three-term difference equation:
\begin{equation}
		E_{n+1}-q_{n}E_{n}+E_{n-1}=
			\lambda_{n+1}-2 u_0\lambda_{n}+\lambda_{n-1},
	\end{equation}
where
	\begin{align}
	E_{n} &=\left[\epsilon Q^{{1}^{\prime}}_{n-\frac{1}{2}}(u_0)-{\cal{P}}_n (u_0)\right] A_{n},\notag\\
			q_{n} &=2 u_0+\frac{(\epsilon-1)Q^{{1}^{\prime}}_{n-\frac{1}{2}}(u_0)}{\epsilon Q^{{1}^{\prime}}_{n-\frac{1}{2}}(u_0)-{\cal{P}}_n (u_0)},\notag\\
				\lambda_{n} &=\left[Q^{{1}^{\prime}}_{n-\frac{1}{2}}(u_0)-{\cal{P}}_n (u_0)\right]E_a(1+\delta_{n0}),
	\end{align}
 which will be solved here using a Green function as described first by Love~\cite{Love1972}, and revisited by Kuyucak, \emph{et al.}~\cite{Kuyucak1998}. 
 Integrating the induced charge density over the volume of the ring using the surface charge, Eq.~\eqref{surfacecharge} yields the induced dipole moment, which is proportional to the polarizability $\alpha_i$. Thus, the dipolar scattering cross section $\sigma_{ds}$ as given by:
\begin{equation}
\sigma_{ds}=\frac{8\pi \omega^4}{3 c^4}\mid\alpha_i\mid^2,
\end{equation} 
can be computed.  
The absorbed power by the ring elevates its temperature $T(\bar{r},t)$ which can be calculated by solving the heat diffusion equation $\rho C_p \dot{T}(\bar{r},t)-k\Delta T(\bar{r},t) = s(\bar{r},t)$, where $\rho$, $C_p $, and $\kappa $ are the density, specific heat, and thermal conductivity of the ring material, respectively. Any modulation information of the amplitude of the incident field will be embedded in $s$.

The solutions above can be used to visualize the quasi-static toroidal fields and the dipolar scattering cross section within the zeroth perturbative order for the two polarization states of the incident photons, and to validate and further facilitate comparison to the computational results for the more accurate time dependent response. 
Prior to obtaining the latter, we first employ the finite element (FE) method ~\cite{Zienkiewicz1977} to solve the Laplace equation in a three dimensional Cartesian computational domain, where an initial uniform static field is established by assigning equal and opposite potential values $\eval {\Psi}_{\pm k}$, $k=x_0,y_0,z_0$  to the appropriate domain boundaries $k$ corresponding to the two considered polarizations $E_{a\parallel}$ and $E_{a\perp}$. We will thus solve $\bar{\nabla} \cdot (\epsilon\bar{\nabla}\Psi) =0$ in the domain containing the torus by applying $\hat{n}\cdot \bar{D} =0$ to all other boundaries. 
Following the quasi-static results,  to obtain the scattering properties of the single as well as several multi-ring configurations, we will solve $\nabla \times \bar{H}= \partial (\epsilon \bar{E})/\partial t + \sigma \bar{E}$ together with $\nabla \times \bar{E}= -\partial  \bar{H}/\partial t $,
employing the finite difference time domain (FDTD) method ~\cite{Kunz1993}. Here, again a Cartesian computational domain with absorbing boundary conditions (perfectly matched layers) is used to mimic the vacuum, and a linearly polarized plane wave is propagated to interact with the torus.

Having determined the response of the  toroidal nanostructure, we will now describe how it may be suspended, and what the effect of such a suspension may be on the electric field of the nanotrap.
For any experimental attempts to trap molecules, the nano-traps need to be suspended without significantly distorting the electric field of the trapping region. As a preliminary suspension method, one may consider a membrane containing single or arrays of toroidal traps. Such an array of nano-traps may be fabricated using a low loss substrate.  Employing a focused ion beam lithography approach, one may attempt to carve channels in the membrane beneath the opening region of the torus, or fabricate the traps on other appropriate meshed substrates. For an optically transparent membrane, we anticipate negligible scattering or modification of the local electric field. Studying the optical responses of metallic nanoparticles, similar configurations arise when using a photon scanning tunneling microscope (PSTM) or a near-field optical microscope (NSOM) for imaging purposes, or for mapping the local field distribution. In such cases, when an optically conductive nanometer sized probe is brought in the proximity of the metal nanostructure (in our case, the torus) to register the field distribution, the question of what effects the approaching probe or the substrate on which the nanoparticle resides may have on the nanoparticle surface modes has arisen in recent investigations. Such effects can be shown to be negligible for non- or low-absorbing substrates, such as fused silica, glass, and for parts of the spectra, silicon and germanium. Although the presented concept is based on an operation away from the plasmon band of the nanostructure and without a strong need for spectral tunability, in plasmon related studies, the (dielectric) substrate effect on the plasmon energy spectrum typically only amounts to a minor redshift. Therefore, substrate effects can be safely assumed to be small, and more importantly, non-detrimental to the trap formation. More specifically, with reference to Eq.\ 3 in ~\cite{Passian2005}, a similar study can be performed to quantify the negligible effect of a low-loss substrate.

\section{Electromagnetic trapping with nano-rings}
\label{sec:3}
Our objective here is to investigate the potential of metal nano-rings in applications related to electromagnetically trapped species possessing a permanent or induced electric dipole moment $\bar{\mu}$.
We wish to show that microscopic objects carrying a moment will be unable to escape from a trapping volume in the vicinity of the central region of the torus.   
We assume that the particles to be trapped are in thermal equilibrium such that their positions and velocities follow an appropriate distribution. Nanoparticles are readily aerosolized 
at room temperature, however, in this study, the assumption is that they, too, have been cooled down and introduced in the toroidal region. Our primary focus here is on trapping molecules.
 The field $\bar{E}$ distribution of the ring is thus to present a potential energy barrier to the molecules. The engendered field by the metal nano-ring perturbs a molecule in a rotational state with an interaction energy of the form $\hat{H}_E=-\bar{\mu}\cdot \bar{E}$ modifying the rotational energy levels of the molecules (Stark perturbation). 
For certain polar molecules used in Stark deceleration studies, the Stark energies of the rotational states have been calculated as a function of the applied field strength~\cite{VandeMeerakker2012}. Specifically, for our objective, the assumptions regarding the quantum states of the molecules are similar to those leading to the Stark energy expression for ammonia isotopologue $^{15}$ND$_3$ (see Eq.\ (11) in ref.~\cite{VandeMeerakker2012}):
\begin{equation}\label{stark}
W_{\text{Stark}}(u,\eta,\phi)=\sqrt{\frac{W^2_{\text{inv}}}{4}+ \left ( \frac{\bar{\nabla}\Psi}{\alpha}\right )^2}-\frac{W_{\text{inv}}}{2},
\end{equation}
where $\Psi$ can be taken from Eq.~\ref{ex} or be provided computationally from FE and FDTD,  $W_{\text{inv}}$ denotes the energy splitting associated with the atomic inversion within the polar molecule in the state $\ket{J,K,M}$, and $\alpha = J(J+1)/\mu K M$ is related to the matrix element diagonal in $K$ and $M$ of the Stark perturbation such that the first order Stark energy is $-E/\alpha = \bra{J,K,M} \hat{H}_E \ket{J,K,M}$. Similarly, second-order Stark energy correction and high field strength Stark effect including higher-order perturbation terms may be considered  as necessary. Using Eq.~\ref{stark} and the field profile of a nanoring with given geometric characteristics, the energy barrier along the $x, y$ and $z$ directions may be studied. 
Furthermore, for the sake of future studies, and without further elaboration here, denoting the mass of the ND$_3$ molecule with $m$, forming the Lagrangian and obtaining  the equations of motion: 
\begin{align}
 \frac{ms^2 \ddot{u}}{(u^2 -1)(u-\cos{\eta})^2}+\frac{\partial W_{\text{Stark}}}{\partial u}&=0\\
  \frac{ms^2 \ddot{\eta}}{(u-\cos{\eta})^2}+\frac{\partial W_{\text{Stark}}}{\partial \eta}&=0\\
   \frac{ms^2 (u^2 -1)\ddot{\phi}}{(u-\cos{\eta})^2}+\frac{\partial W_{\text{Stark}}}{\partial \phi}&=0,
\end{align}
the trapped molecular  motion may be studied within the trap volume or molecular deceleration may be investigated for molecules moving through the trapping region. 

\begin{figure}
	\centering
	\includegraphics[width=8cm]{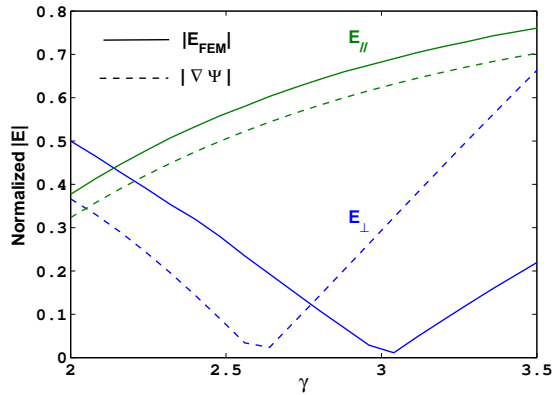}
	\caption{\label{fig:3}Polarization dependence of the field at the center of the torus. For a gold nano-ring of appropriate aspect ratio $\gamma$, when the polarization of the incident field is perpendicular to the symmetry axis of the torus, the electric field can be made negligibly small. The analytical solution using first order perturbation, while predicting a lower value for the aspect ratio for this minimal field, overall it agrees well with the computational solution employing the finite elements method.}
\end{figure}
\begin{figure}
	\includegraphics[width=9cm]{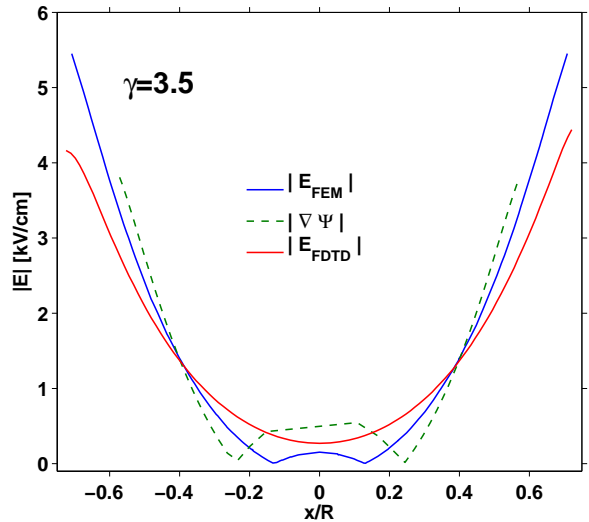}
	\caption{\label{fig:4} Comparison of the quasi static and time dependent electric field across the torus.  For an aspect ratio $\gamma = 3.5$, the analytical results as well as  the computational results in both the electrostatic (FE) and transient (FDTD) limits consistently produce a field profile suggestive of a trap formation. The field lines computed using the FDTD method are noted to display smooth curvature, whereas both the analytic and quasi-static computational approaches produce minor artifacts around the low filed regions of the trap.} 
\end{figure}
\begin{figure}
\centering
 \includegraphics[width=9cm]{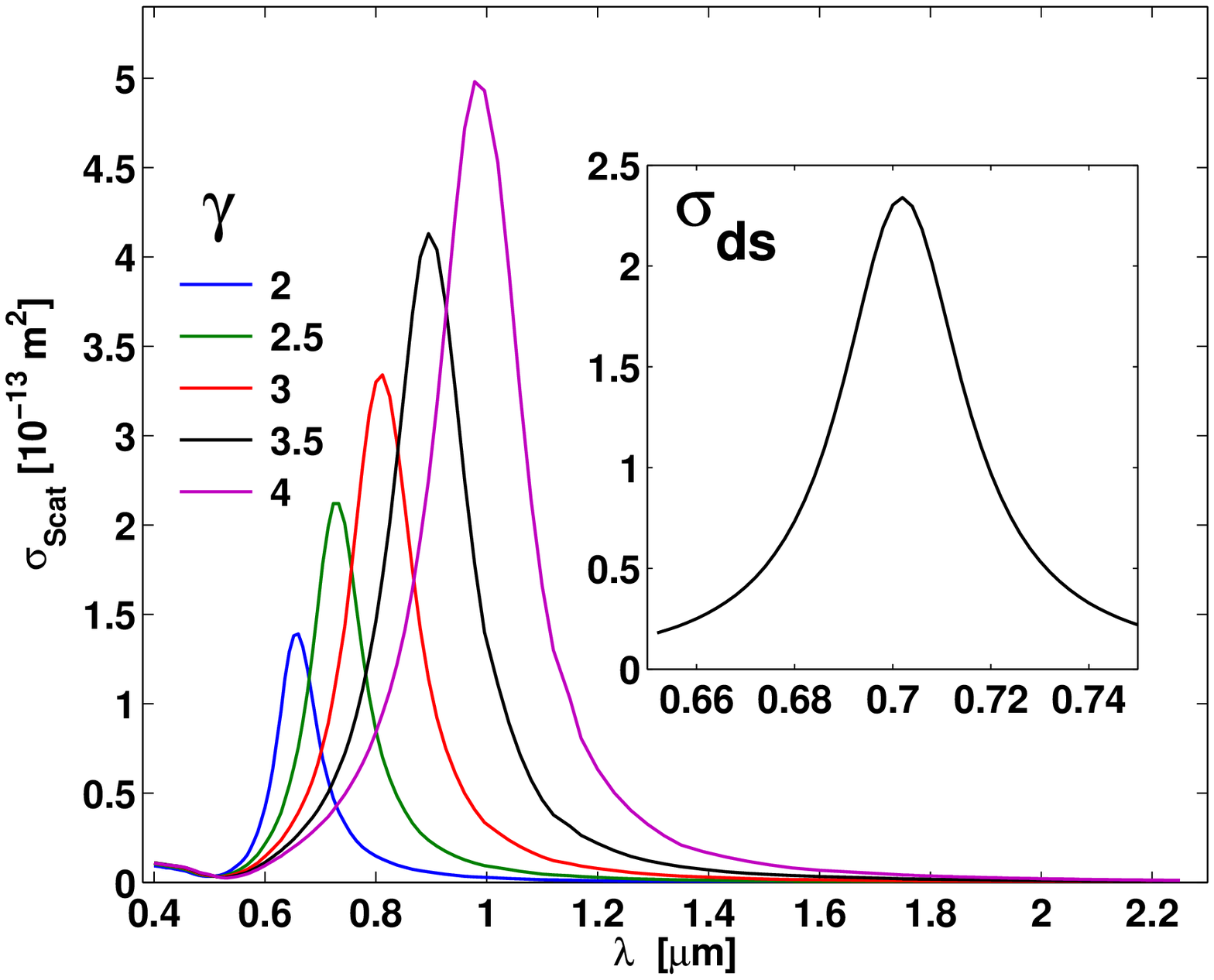}
  \includegraphics[width=9cm]{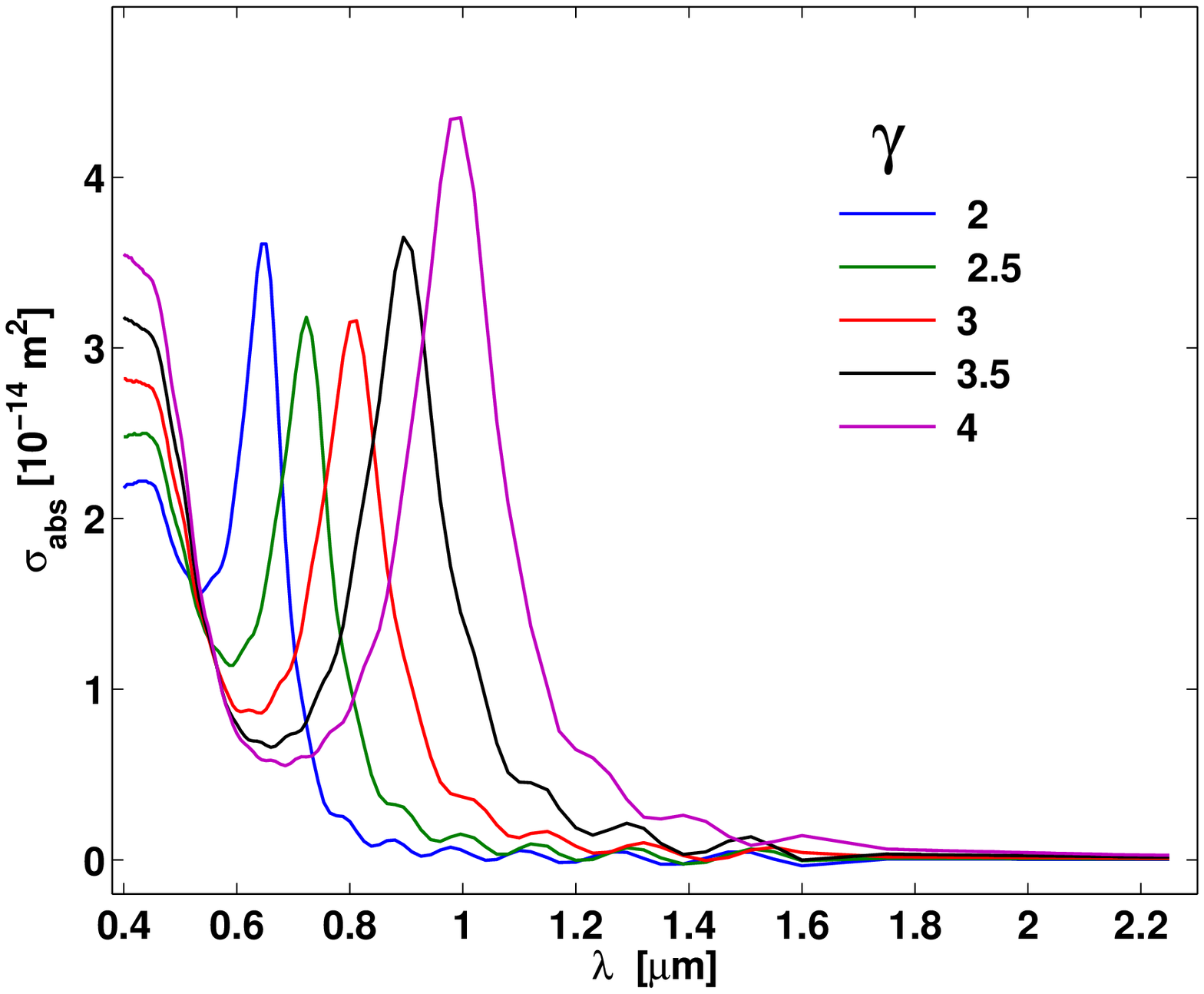}
  \caption{\label{fig:2}  Computed scattering (upper panel) and absorption (lower panel) cross sections for single gold nanorings of different aspect ratios $\gamma$. For a fixed minor radius $r$, the larger aspect ratios exhibiting the larger cross sections correspond to rings with larger main radii $R$. 
}
\end{figure}  
\begin{figure}
\includegraphics[width=8.5cm]{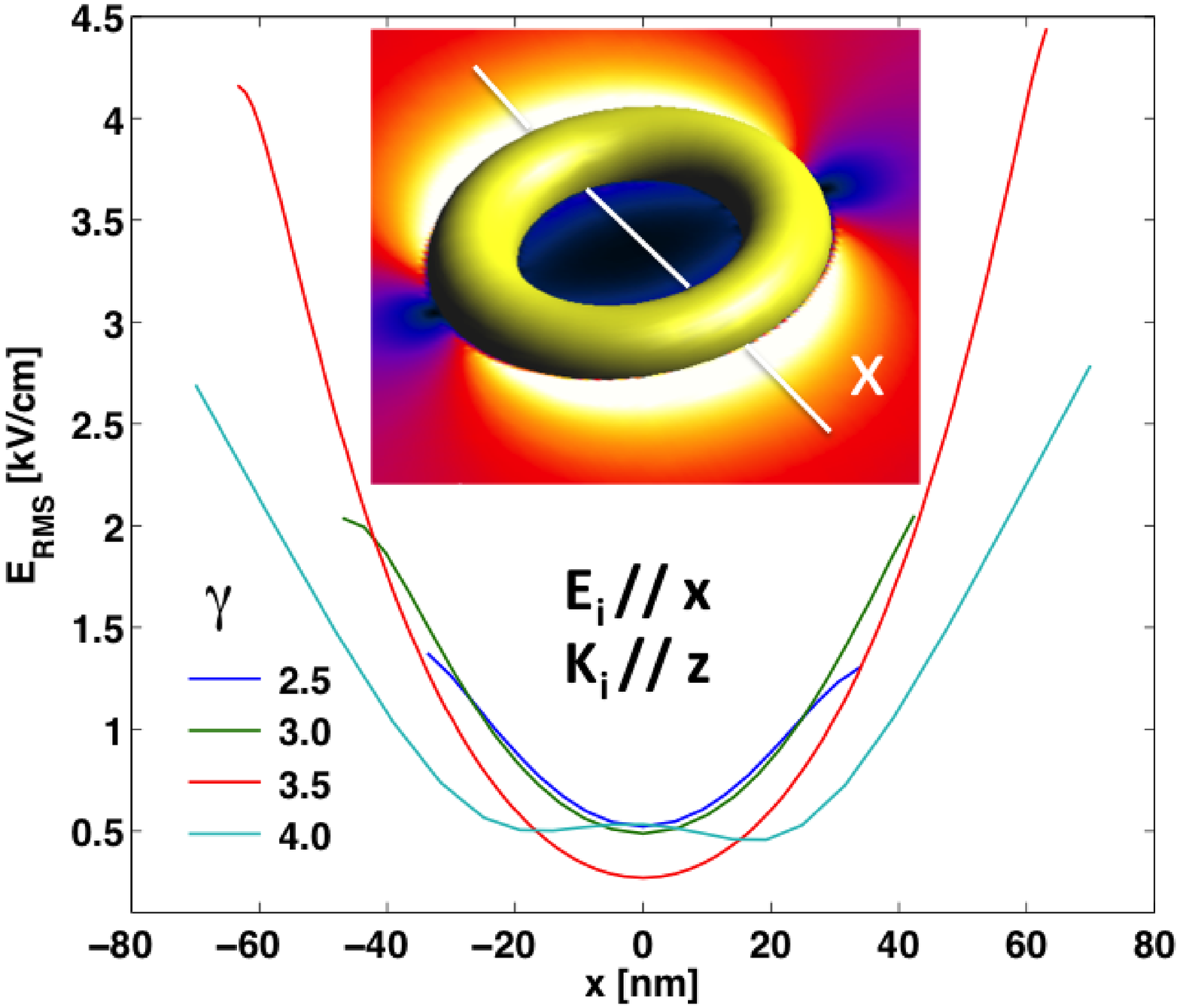}
     \includegraphics[width=8.5cm]{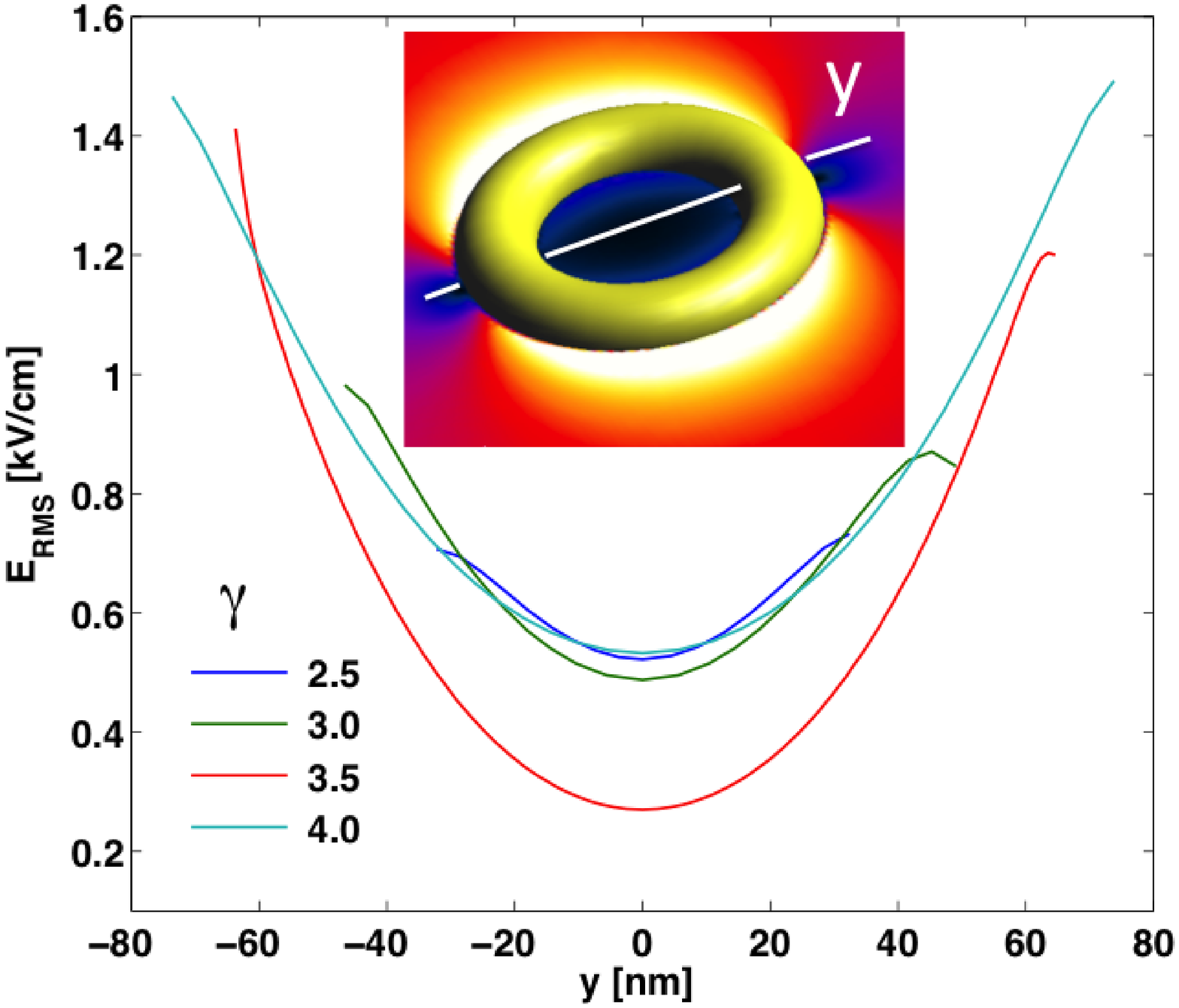}
  \includegraphics[width=8.5cm]{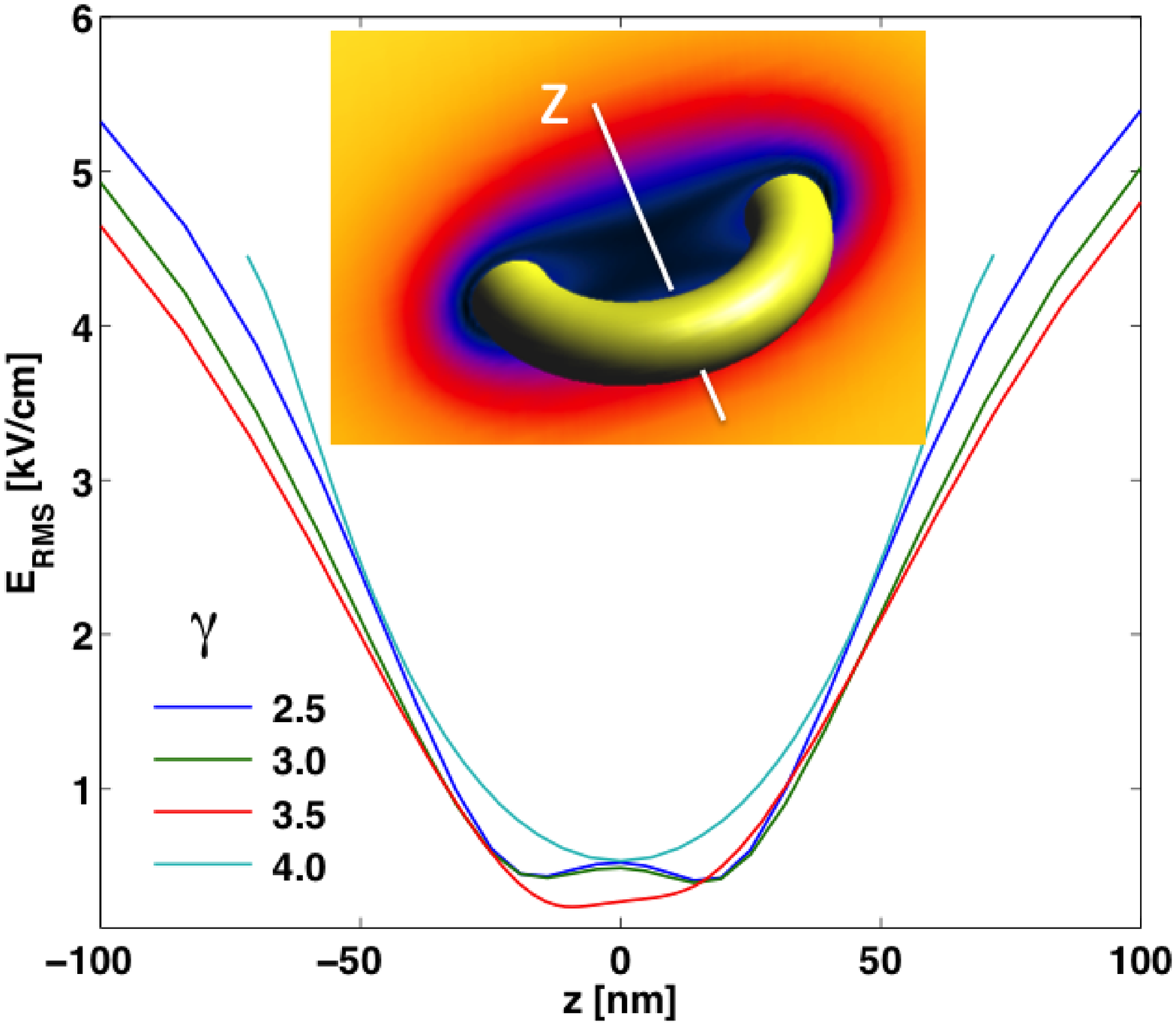}
\caption{\label{fig:5}  Field profiles along $x$, $y$, and $z$ directions across the torus for various aspect ratios for an incident field $E_i$ polarized perpendicular to the symmetry axis $z$ of the torus. The insets visualize the field distributions in the corresponding planes for a ring with $\gamma =3.5 $ interacting with an incident field of a wavelength $\lambda =2 ~\mu m$ propagating along $z$. The field distributions were obtained from FDTD computation.}
\end{figure}
\begin{figure}
\centering
 \includegraphics[width=9cm]{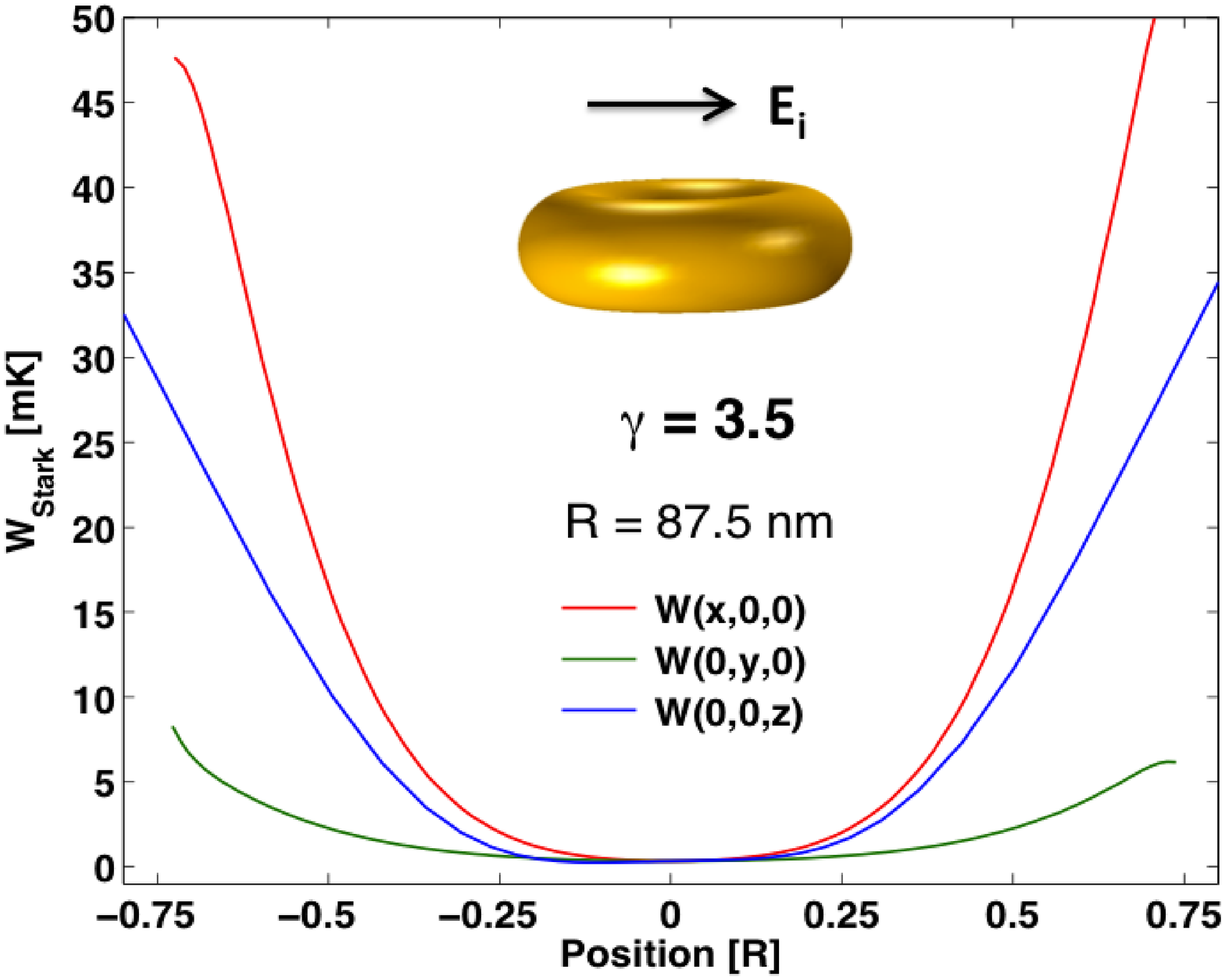}
  \includegraphics[width=9cm]{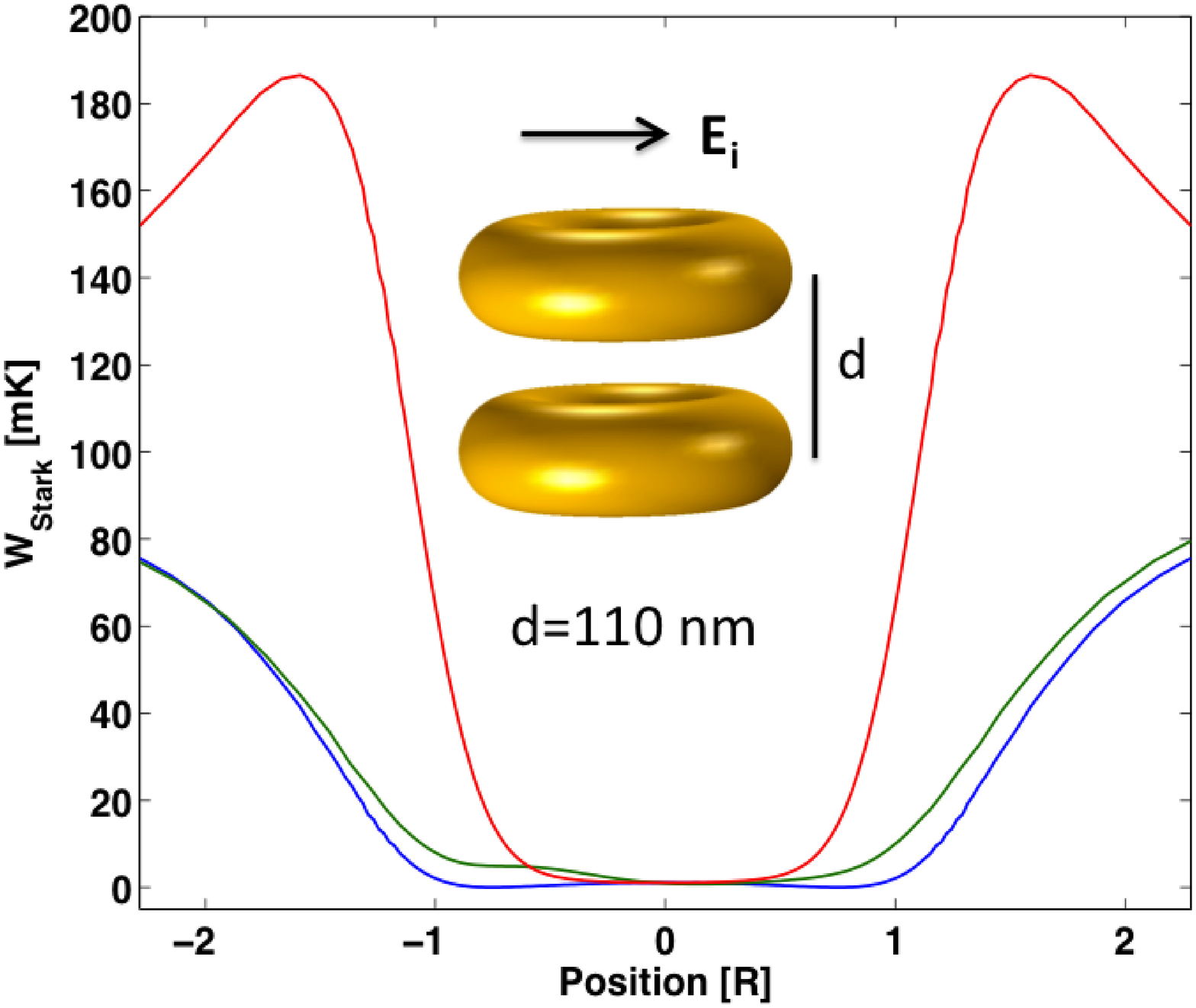}
  \caption{\label{fig:6}   (a) Stark energy profiles along $x$, $y$, and $z$ directions through the center of the torus corresponding to the energy of $ND_3$ molecules interacting with the total toroidal  field, that is, the incident field $E_i$ polarized in the plane of the torus and the scattered field. The field distribution was obtained from FDTD computation. (b) Stark energy profiles through the center of an vertical dimer with constituent rings separated a distance $d= 110$~nm and having an aspect ratio $\gamma=3.5$.}
\end{figure}

 \section{results and discussion} \label{sec:4}

We begin our study of the trapping properties of metal tori by characterizing the optical response of single gold ring tori in the spectral range from visible to the short wavelength infrared. We will then extend the results by considering more elaborate configurations of dimers and trimers as building blocks for arrays of nanorings for both trapping and deceleration of cold polar molecules. To study the geometric dependence of the trapping potential, we introduce  $\gamma =R/r = u_0$, the aspect ratio, where $R= au_0/\sqrt{u_0^2 -1}$ is the major radius and $r = a/\sqrt{u_0^2 -1}$ is the cross sectional radius.  
The material making up the torus is taken to be gold, although other suitable materials may be considered. To determine the field distribution and energy dissipation by the proposed structures, we will resort to simulations, based on the analytical results, and computation employing the FE and the FDTD methods. For the latter, the Drude model is numerically adjusted to fit the values of the complex dielectric function of gold, $\epsilon$, reported by Johnson and Christy~\cite{Johnson1972}. The fitting parameters $\omega_D=2155.6$~THz, $\gamma_D=18.36$~THz and $\epsilon_\infty=9.06$ reproduce the values in the spectral range $0.4-2.2$~$\mu m$, which will be considered for the FDTD simulation of the electric field and absorption spectra of the various ring configurations studied. 

The results of both the FE computation and the perturbative analytic calculation for the field value at the center of the ring and its dependence on the aspect ratio are plotted in Fig.~\ref{fig:3}. The results, being in good agreement, suggest a special value of the aspect ratio $(\gamma \approx 3)$ in the case of perpendicular polarization of the incident beam  $(\bar{E_a} \perp  \hat{z})$ for which the field is almost extinct.  Further analysis sets this values closer to $\gamma \approx 3.5$, for which we then compare the results of the FE, FDTD and perturbative analytic calculations by plotting the field along the polarization direction for each method in Fig.~\ref{fig:4}. Evidently, for the geometric configuration at the given aspect ratio, the field distribution suggests the possibility  of forming a trap.

In the time domain, the interaction of either a continuous wave (CW) or an amplitude modulated (AM)  laser beam with the torus results in a total absorbed power $P_t$. Assuming that a linearly polarized field $\bar{E}(\bar{r},t)$ of wavelength $\lambda$ and propagation vector $\bar{k}$ can be provided by a an appropriate laser, we specify the incident radiation to be in the form of a beam of a circular cross sectional area $A$,  pulse duration $\Delta t$,  the repetition rate $T_l=1/f_l$, energy per pulse $e_p$ and an average power $P_a$. 
For a single gold nanoring with an aspect ratio $\gamma$ and volume $V$, denoting the incident intensity $I_i$, the absorption and scattering cross sections are defined as $\sigma_{abs} = \int_V pdV/I_i$ and $\sigma_{scat} =\int_S \bar{S}_{scat}\cdot \bar{dS}/I_i$, respectively, where $p$ is the absorbed power density in the ring and $S_{scat}$ is the scattered Poynting vector over its surface. Neglecting photoluminescence, the extinction cross section is  $\sigma_e = \sigma_{abs} + \sigma_{scat} $ and the total absorbed power by the particle is given by $P_t = \sigma_{abs} I_i$. Obtaining the scattering and absorption cross sections is relevant to our discussion, as it completes our understanding of the optical response of the nanoring to an incident wave. The response is critical to the feasibility of the nano-trap. Scattering and absorption cross-sections are plotted for different aspect ratios and wavelength range $\lambda \in \left[0.4,2.2\right]~\mu$m in Fig.~\ref{fig:2}. The electromagnetic trap that we are discussing is illustrated by plotting the field profile in the vicinity of the nanoring along the $x, y$ and $z$ directions for different aspect ratios in Fig.~\ref{fig:5}.

To describe the trapping of  cold polar molecules, we will consider ND$_3$, where the molecular beam is in a single weak field seeking state $|J,K,M\rangle = |1,1,-1\rangle$. The ND$_3$ system is known to have an electric dipole moment of $\mu = 1.48$~D and a frequency $W_{\text{inv}}=1.43$~GHz at which the position of the nitrogen atom inverts~\cite{VandeMeerakker2012}). To minimize escape from the trapping potential, the molecular population can be prepared to have appropriate mean-free-path so as to render molecule-molecule encounters less likely. For the trapping volumes considered here,  this can be readily achieved already in the low vacuum regime. 
Assuming that when the applied electric  field is switched off, that is, when $E = 0$, 
the energy of a free cold molecule of mass $m$ and temperature $T$ can be sampled from a Maxwellian distribution, it will have a root mean squared speed of $v= (3k_BT/m)^{1/2}$~m/s, where $k_B$ is Boltzmann's constant. For $T= 25$~mK~\cite{Bethlem2002a} a velocity of $v= 6$~ms$^{-1}$ implies that a cold molecule may depart from the trapping region of a ring with $\gamma =3.5 $ and $R=87.5$~nm. When $E\ne 0$, a threshold trapping energy $W_{\text{Stark}}^{th}$ in Eq.~\eqref{stark} corresponds to a minimum field magnitude of $E^{th} = \alpha [W_{\text{Stark}}^{th}   (W_{\text{Stark}}^{th} + W_{\text{inv}}) ]^{1/2}$. 
From deceleration studies by  Bethlem, \emph{et al.}~\cite{Bethlem2002a},  $W_{\text{Stark}}^{th}>25$~mK, yielding a minimum field amplitude of 
$E^{th}>0.27$~V/$\mu$m. 
Such a value for the local electric field, which can be estimated to be of the same order of magnitude as the incident field due to the spectral position of the plasmon band being far from the laser wavelength, can be made available from an appropriate continuous wave  (CW) laser.
We also note that high resolution infrared absorption spectra of $^{15}$ND$_3$ has been reported by Can\`e, \emph{et al.}~\cite{Cane2013} and thus the incident laser wavelength can be tuned accordingly so as to minimize excitation of the molecular bands.

Furthermore, for a given field amplitude $E^{th}$, the total absorbed power $P_t = \frac{1}{2} c n \epsilon_0 \sigma_{abs} |E^{th}|^2 $ will be below the melting point of the nanostructure.
Since for the typical rings considered here the volume is $V \approx 1.1 \times 10^{-15}$~cm$^3$, we assume that the deposited energy is distributed uniformly, \emph{i.e.}, $s(\bar{r},t) = s(t) =P_t  t $ is the source term obtained from the total absorbed power by the nanostructure.
Employing FE method, we here computationally verified the temperature uniformity by solving for the time dependent thermal response of a gold nanoring with a heat capacity $C_p=129$~J K$^{-1}$kg$^{-1}$, a thermal conductivity $\kappa=317$~Wm$^{-1}$K$^{-1}$, and a density $\rho=19300$~kg m$^{-3}$. The results also predict negligible temperature increase due to the absorbed power of a silicon substrate-supported gold ring in vacuum but otherwise subject to mK near substrate boundary temperature. 

\begin{figure}
 	\centering
 	\includegraphics[width=4.25cm]{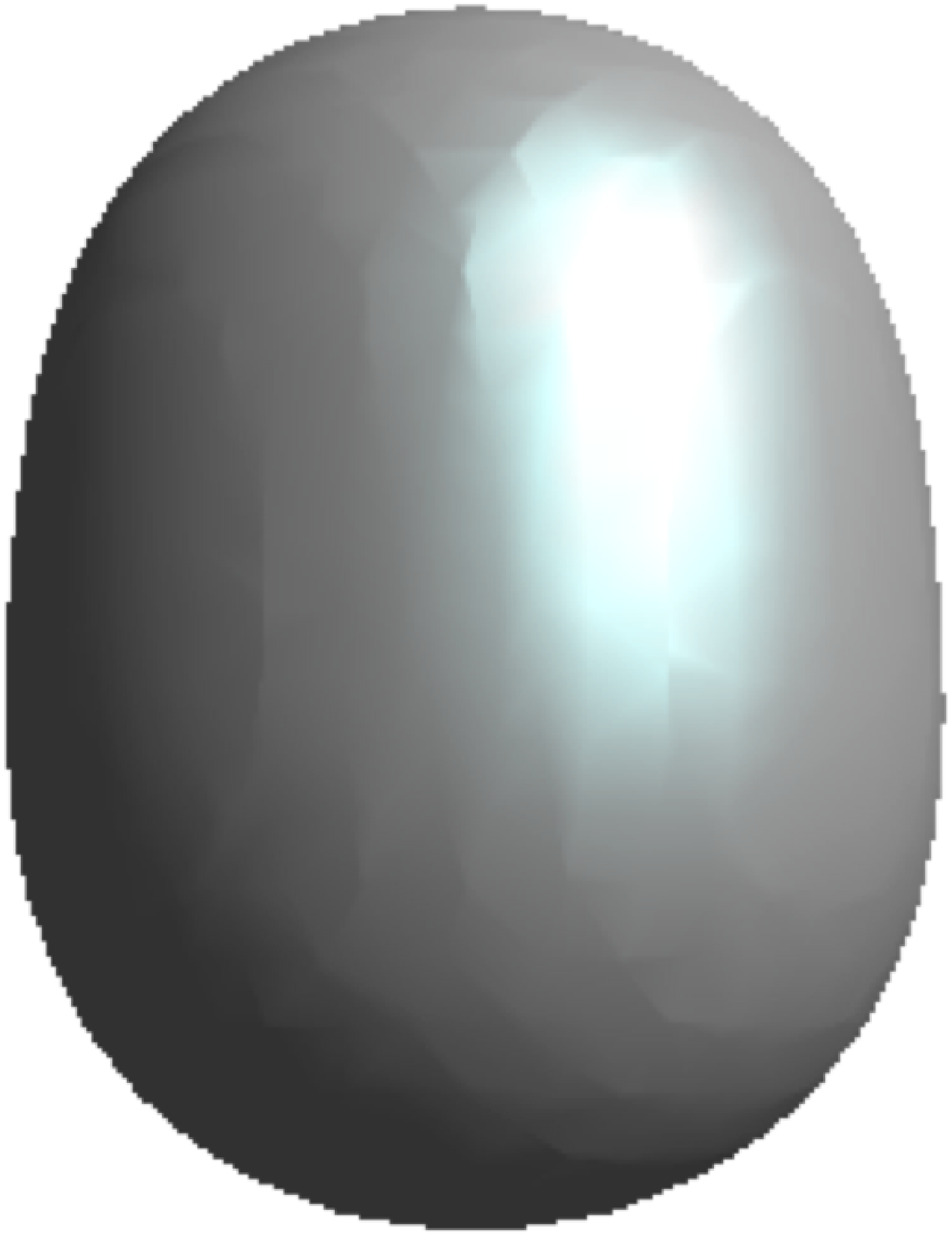}
	\includegraphics[width=4.25cm]{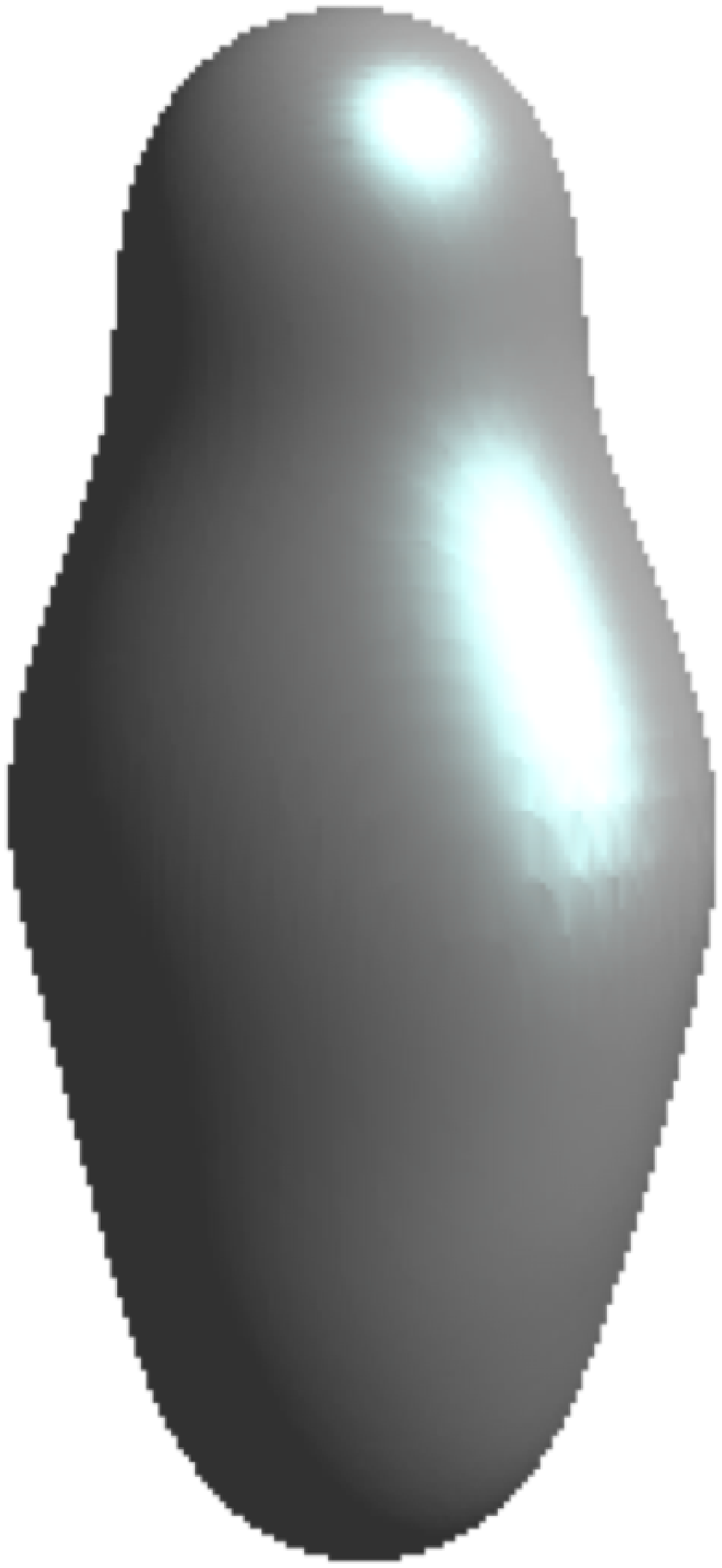}
 	\caption{\label{fig:vol}Visualizing trapping boundaries for the  $^{15}$ND$_3$ system under circular polarization excitation. 
The field isosurface rendering a single ring trapping volume (left) of $2.7\times 10^5$~nm$^3$ and a ring dimer trapping volume  (right) of approximately $5.5 \times 10^5$~nm$^3$. At these boundaries, molecules in a single weak field seeking state $|J,K,M\rangle = |1,1,-1\rangle$ would possess a Stark energy of $W_{\text{Stark}} = 10$~ mK.}
 \end{figure}

 Using Eq.~\eqref{stark}, the stark energy and the field profile of nano-rings with an aspect ratio $\gamma=3.5$, we plot the corresponding energy barrier along the $x, y$ and $z$ directions for two systems, with the choice of the aspect ratio based on the depth of the potential energy barrier. The first system is a single gold nano-ring, whereas the second system is made of two axially aligned gold nano-rings with an optimized separation distance $d=110$~nm for higher energy barrier as illustrated in Fig.~\ref{fig:6}. The dimer ring trapping system has a higher energy barrier when compared to the single ring system. The latter is a realistic trap for molecules with a translational energy along the $z$ direction of the order of few~mK.  
 The dimer trap frequencies reaching the GHz range may contain several tens of bound levels (about 75 at about 1 GHz~\cite{blumel2012}).
To compare the actual trapping volume in both cases, we note that both the position and shape of the  trapping boundary can be affected by the specific $W_{\text{Stark}}^{th}$ and the incident photon polarization, as shown in Fig.~\ref{fig:vol} for the case of circular polarization.
 \begin{figure}
 	\centering
 	\includegraphics[width=9cm]{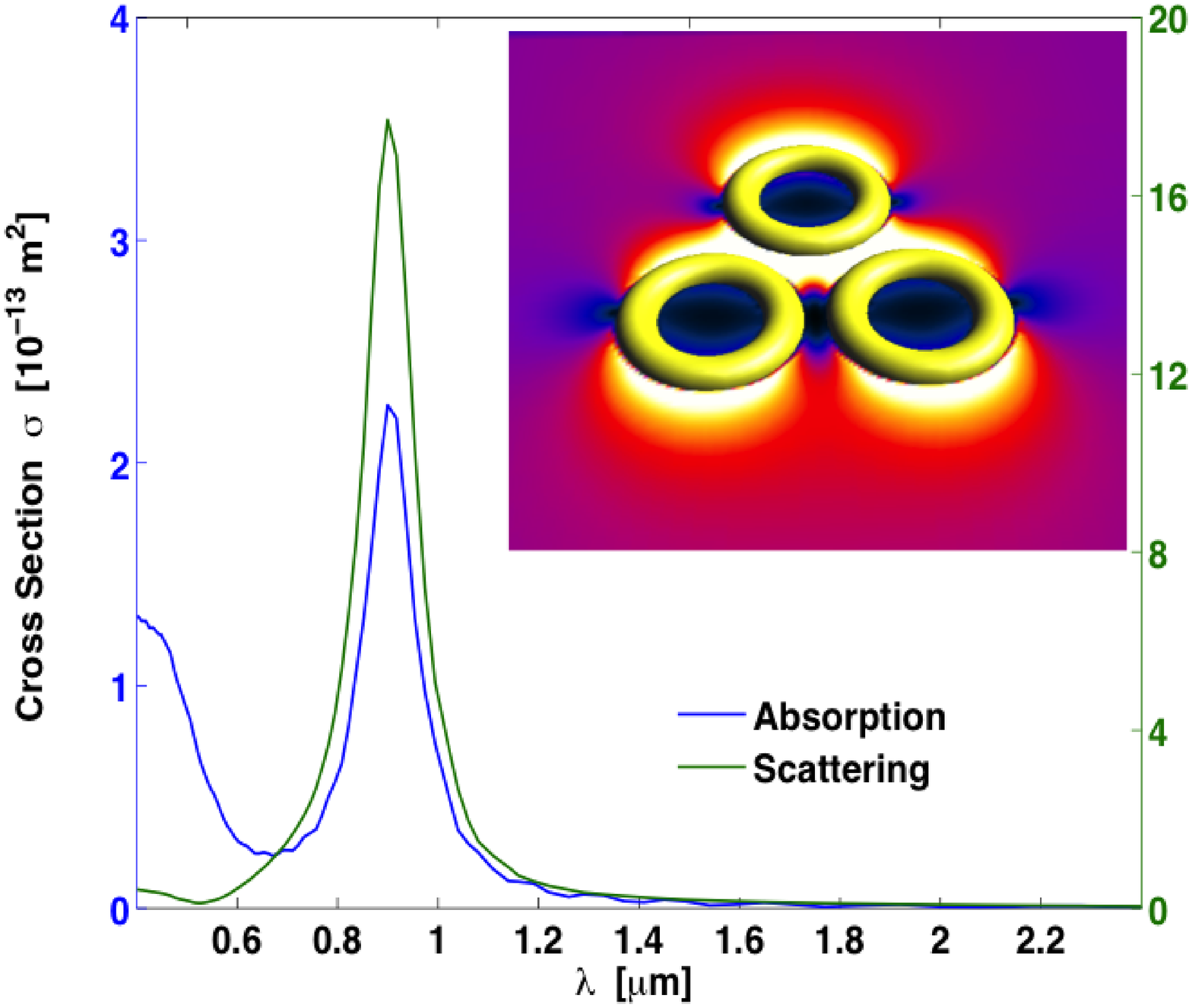}
 	\includegraphics[width=9cm]{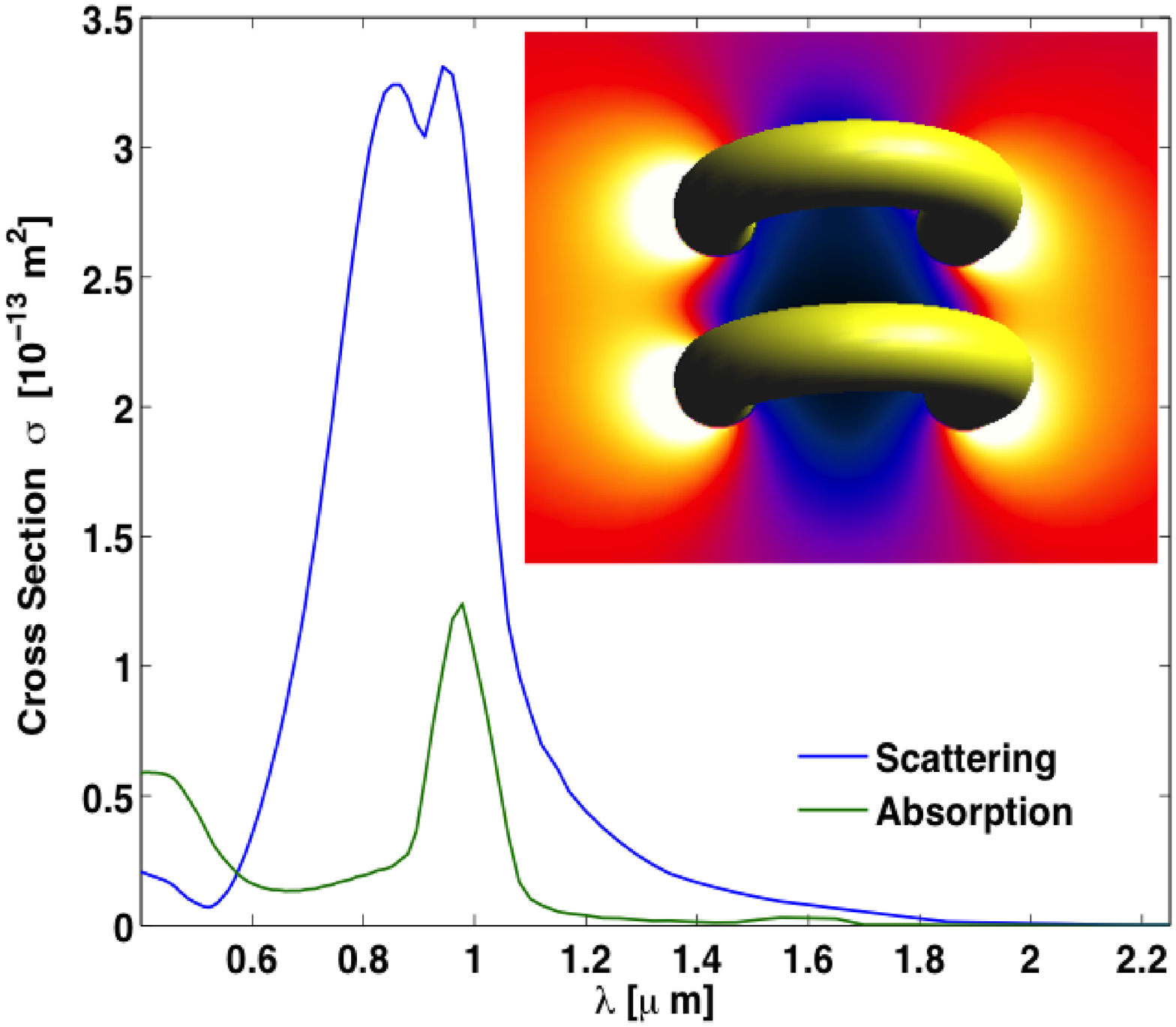}
 	\caption{\label{fig:7} (a) Computed scattering and absorption cross sections for single planar gold nanoring trimer  with each ring having  $\gamma$. With the rings sufficiently apart, the cross sections approach those of a single ring.
 		(b) Comparison of the absorption and scattering cross sections for a vertical dimer with the same parameters as those in Fig.~\ref{fig:5}.}
 \end{figure}  
  
The proposed concept can be extended to include another important case where the toroidal nano-trap may be considered as a potential building block of a quantum computer, which requires the manipulation of qubits, as discussed, e.g., by DiVincenzo~\cite{DiVincenzo2000}. The most elusive requirement for building a quantum computer is to have a set of individually addressable qubits. This translates into the formation of a 2D array of nano-rings. The array is characterized by a  minimum separation distance that would preserve the electromagnetic trap profile, found in our example simulation to be $d_{min} \approx 0.5~\mu$m between individual centers of the rings. Thus, to verify that the individual rings are not coupled in a way that would alter the trap profile, we obtain the absorption and scattering cross section for an array of rings with a minimum separation $d_{min}$. Both plots are shown in Fig.~\ref{fig:7}. One can easily see that they are almost identical to the absorption and scattering plots obtained in the case of individual rings.

In closing, we note that an experimental exploration of the proposed concept may begin in a configuration of scanning probe microscopy. For example, the hybrid photonic-nanomechanical force microscopy (HPFM) or the near-field scanning optical microscope (NSOM), both with spatial resolutions below the diffraction limit and 
allowing high levels of control and manipulation of light at the nanoscale, may be employed to first characterize the electromagnetic response and morphology of single nanorings. Similar modalities permit spectroscopic interrogation by both delivering the excitation photons locally to the trapped molecules as well as collecting the emitted photons by the molecules.

\section{Conclusion}\label{sec:5}
The theoretically obtained scattering properties and field distributions of toroidal gold nano-particles strongly suggest that trapping of cold polar molecules and dielectric beads may be feasible optically in the proposed geometry. For a wavelength of 2~$\mu$m, the results for a specific value of the toroidal aspect ratio indicate the possibility of a strong and well defined trap for cold polar molecules when the external field polarization is parallel to the axis of the ring. Interestingly, the results show that a trap forms away from plasmonic resonances of the nano-ring. The frequency dependence of the nano-ring response shows an enhancement of the electric field near the plasmon resonance band, as may be expected. However, at resonance the trap becomes shallow.  Furthermore, from the single torus results, we suggest 
more elaborate configurations in the form of  two-dimensional arrays of nano-rings toward the potential implementation of multi-particle nano-traps. Such compact and miniaturized structures may be of importance in the  design of future hardware platforms for quantum computation applications.
Considering the large forces on the molecules achieved in the nanotrap with trap frequencies up to a GHz, it is conceivable to resolve these levels spectroscopically.
The utility of the toroidal trapping concept may be extended to confining dielectric particulates via force gradients similar to dielectrophoresis~\cite{Pohl1978},
by creating a toroidal potential energy barrier.
Assuming a particle with no internal degrees of freedom but with a dielectric function $\epsilon_p$, the force leading to stable small amplitude oscillations of the particle may be obtained from Maxwell's stress tensor $\mathbf{T}$ by integration over the particle surface $\bar{f}(t)= \iint \mathbf{T} \cdot \hat{n}dA$. 
Thus, particles such as polystyrene nano-spheres may be trapped via a similar approach.

\acknowledgments
This research was supported in part by the laboratory directed research and development fund at Oak Ridge National Laboratory (ORNL). ORNL is managed by UT- Battelle, LLC, for the US DOE under contract DE-AC05- 00OR22725.

\bibliographystyle{kp}
 \bibliography{traprefs}
  
\end{document}